\documentclass[useAMS,usenatbib]{mn2e}
\usepackage{amsmath,amssymb,graphicx}
\usepackage{subfigure}
\def\deg{\hbox{$^\circ$}}              
\def\arcm{\hbox{$^\prime$}}            

\def\p{\hbox{$P$}}              
\def\pa{\hbox{$\theta$}}              
\def\pav{\hbox{$\theta_V$}}              
\def\pv{\hbox{$P_V$}}              
\def\pr{\hbox{$P_R$}}              
\def\pb{\hbox{$P_B$}}              
\def\Pab{\hbox{$\theta_B$}}
\def\Pav{\hbox{$\theta_V$}}
\def\Par{\hbox{$\theta_R$}}

\title[LO Peg: Optical polarization]{LO Pegasi: An investigation of multi-band optical polarization}
\author[J. C. Pandey et al.]{J. C. Pandey\thanks{jeewan@aries.res.in(JCP), biman@aries.res.in(BJM), sagar@aries.res.in(RS), pandey@aries.res.in(AKP)},  Biman. J. Medhi$^*$, R. Sagar$^*$, and A. K. Pandey$^*$ \\
Aryabhatta Research Institute of Observational Sciences, Nainital - 263 129, India}

\date{Accepted 2009 March 10.  Received 2009 March 4; in original form 2008 May 30}

\begin{document}

\pagerange{\pageref{firstpage}--\pageref{lastpage}} \pubyear{2008}

\maketitle   \label{firstpage}

\begin{abstract}
We present  BVR polarimetric study of the cool active star LO Peg for the first time.  LO Peg was found to be highly polarized among the cool active stars. Our observations yield average values of polarization in LO Peg:  $P_B =0.387\pm 0.004\%$, $\theta_B = 88\deg\pm1\deg$; $P_V=0.351\pm0.004\%$, $\theta_V=91\deg\pm1\deg$; and $P_R= 0.335\pm0.003\%$, $\theta_R = 91\deg\pm1\deg$.  Both the degree of polarization and the position angle are found to be variable. The semi-amplitude of the polarization  variability in B, V and R bands are found to be $0.18\pm0.02$\%, $0.13\pm0.01$\% and $0.10\pm0.02$\%, respectively.
We suggest that the levels of polarization observed in LO Peg could be the result of scattering of an anisotropic stellar radiation field by an optically thin circumstellar envelope or scattering of the stellar radiation by prominence-like structures.
\end{abstract}


\section{Introduction}
\label{sec:intro}
Magnetically active late-type stars have inhomogeneities on their surfaces that may cause variation in their spectral lines and light curves.  The inhomogeneous distribution of magnetic regions on the surface of late-type dwarfs may also produce a broad-band linear polarization (Tinbergen \& Zwaan 1981), and it is expected that the polarization of the integrated stellar light may change along with stellar activity phenomenon (e.g. Huovelin et al. 1989; Mekkaden et al. 2007).  Several active stars  have small, variable amounts of linear polarization ($<0.1\%$) at optical wavelengths that are  best interpreted as a result of scattering from cool circumstellar material (Scaltriti et al. 1993).  Liu \& Tan (1987) have shown that for most of RS CVn binaries the optical polarization is weak, generally below 0.45\%, averaging about 0.2\%. Pfeiffer (1979) found 0.32\% polarization in proto-type RS CVn and concluded that the polarization was due to scattering from cool, transient, circumstellar material probably ejected in the clouds. In BH CVn the polarization in U band was found to vary synchronously with the binary period  with a peak-to-peak amplitude of 0.03\%. It was attributed to the reflection of light coming from  bright primary by the envelope of secondary (Barbour \& Camp 1981). In the late-type stars linear polarization may be due to inhomogeneities  or non-isotropic gas flow. The inhomogeneities is mainly due to the  magnetic areas (star spots or plages) and circumstellar gas or dust envelopes. The most popular mechanisms that have been suggested as a source of linear polarization in late-type stars are Rayleigh, Mei and Thomson scattering in an optically thin medium (e.g Brown \& McLean 1977; Brown, Mclean \& Emslie 1978; Pfeiffer 1979; Piirola \& Vilhu, 1982; Yudin \& Evans 2002) and magnetic intensification (e.g. Leroy 1962; Mullan \& Bell 1976; Huovelin \& Saar 1991; Saar \& Houvelin 1993).

LO Peg  is a single, young, K3V-K7V type and  a member of the Local Association (Jeffries \& Jewell 1993; Montes et al. 2001; Gray et al. 2003; Pandey et al. 2005). It is one of the fast rotating active star with a period  of 0.42 day.  LO Peg shows strong H$\alpha$ and Ca {\sc II} H and K emission lines (Jeffries et al. 1994).  Evidence of an intense downflow of material and optical flaring on LO Peg have been presented by Eibe et al.  (1999). Recently, Zuckerman, Song \& Bessell  (2004) have identified LO Peg as a member of a group of 50 Myr old stars that partially surround the Sun. Optical photometric  light curves and related studies of LO Peg have been carried out by Pandey et al. (2005). Its relation with polarimetric properties is still not known and the same has been investigated here for the first time.  The observations, the methods of data reduction, results and discussion  along with the  conclusions are given in forthcoming sections.

\section{Observations}
\label{sec:obs}
The broad band $B$ ($\lambda_{eff}$=0.44  $\mu m$), $V$ ($\lambda_{eff}$=0.55 $\mu m$)  and $R$ ($\lambda_{eff}$=0.66  $\mu  m$) polarimetric observations of LO Peg  have been made in between  October 19 to  December 19, 2007 using ARIES Imaging Polarimeter (AIMPOL; Rautela, Joshi \& Pandey  2004; Medhi et al. 2007), mounted on the Cassegrain focus of the 104-cm Sampurnanand  telescope (ST) of ARIES, Nainital.  The imaging was done  by TK $1024\times 1024$ pixel$^2$  CCD camera.  Each square pixel of the CCD corresponds to $1.73$ arcsec while the entire field of view of CCD is $\sim 8$  arc-min in diameter on the  sky.  The read out noise and gain of the CCD  are 7.0 $e^-$  and 11.98 $e^-$/ADU, respectively.  For linear polarimetry the retarder (Half-wave plate) is rotated at 22.5$^\circ$ intervals between exposures. Therefore, one polarization measurement was obtained from every four exposures (i.e. at the retarder position of $0^\circ$, $22^\circ.5$, $45^\circ$ and $67^\circ.5$).  The ordinary and extraordinary images of each source in the CCD frame are separated by 27 pixels along the north-south direction on the sky plane.  Standard CCD procedures in {\sc iraf}\footnote{{\sc iraf} is distributed by the National Optical Astronomy Observatory} (bias subtraction, centroid determination) were applied to extract  the flux of two stellar images in each CCD frame.  Following relation was used to obtained the degree of polarization (\p) and polarization potion angle (\pa).

\begin{equation}
R(\alpha) = \frac{I_O/I_E -1}{I_O/I_E+1} = P cos(2\theta - 4 \alpha)
\end{equation}

\noindent
which is the difference between the intensities of the ordinary ($I_O$ ) and extraordinary ($I_E$ ) beams to their sum. Here, $\alpha$ is angle which the retarder makes with north-south direction.  The detail descriptions about the  AIMPOL, data reduction and calculations of polarization, position angle are given in Rautela, Joshi \& Pandey (2004) and Medhi et al. (2007).

\begin{table*}
\centering
\caption{Observed polarized and unpolarized standard stars.}\label{std_obs}
\begin{tabular}{ccccccc}
\hline
\multicolumn{5}{|c|}{Polarized Standard}&\multicolumn{2}{c}{Unpolarized Standard}\\
\hline
& \multicolumn{2}{c}{Schmidt et. al    (1992)}&\multicolumn{2}{c}{This work}&\multicolumn{2}{c}{This work}\\
\hline
Filter&$P           (\%)$ &  $\theta             (^\circ)$ &  $P           (\%)$ & $\theta            (^\circ)$ &\ \ \ \ $q(\%)$ &\ \ \ \ \ $u(\%)$ \\
\hline
\multicolumn{5}{|c|}{\underline{HD 25433}}                               &\multicolumn{2}{|c|}{\underline{HD21447}} \\
B & $5.23\pm0.09$ & $134.3\pm 0.5 $  & $5.17\pm 0.21$&$ 135.1  \pm 1.0 $ & \ \ \ \ \ 0.019  &\ \ \ \ \ \ 0.011  \\
V & $5.12\pm0.06$ & $134.2\pm 0.3 $  & $5.13\pm 0.09$&$ 134.7  \pm 0.8 $ & \ \ \ \ \ 0.037  &\ \ \ \ -\  0.031 \\
R & $4.73\pm0.05$ & $133.6\pm 0.3 $  & $4.76\pm 0.13$&$ 132.9  \pm 0.5 $ & \ \ \  -\ 0.035  &\ \ \ \ -\  0.039 \\
\multicolumn{5}{|c|}{\underline{HD 204827}}                              &\multicolumn{2}{|c|}{\underline{HD12021}} \\
B & $5.65\pm0.02$ & $58.20\pm 0.11$  & $5.72\pm 0.09$&$  58.6 \pm 0.5$ & \ \ \  -\ 0.108  & \ \ \ \ \ \ 0.071 \\
V & $5.32\pm0.02$ & $58.73\pm 0.08$  & $5.35\pm 0.03$&$  60.1 \pm 0.2$ & \ \ \ \ \ 0.042 & \ \ \ \ -\   0.045 \\
R & $4.89\pm0.03$ & $59.10\pm 0.17$  & $4.91\pm 0.20$&$  58.9 \pm 1.2$ & \ \ \ \ \ 0.020 & \ \ \ \ \ \ 0.031 \\
\multicolumn{5}{|c|}{\underline{BD+59$^{\circ}$389}}                     &\multicolumn{2}{|c|}{\underline{HD14069}} \\
B & $6.35\pm0.04$ & $98.14\pm 0.16$  & $6.36\pm 0.13$&$  97.5 \pm 0.5$ &\ \ \ \ \ 0.138 &\ \ \ \  -\ 0.010 \\
V & $6.70\pm0.02$ & $98.09\pm 0.07$  & $6.80\pm 0.07$&$  98.2  \pm 0.2$ &\ \ \ \ \ 0.021 &\ \ \ \ \ \ 0.018 \\
R & $6.43\pm0.02$ & $98.15\pm 0.10$  & $6.39\pm 0.04$&$  99.5  \pm 0.2$ &\ \ \ \ \ 0.010 &\ \ \ \  -\ 0.014 \\
\multicolumn{5}{|c|}{\underline{HD 19820}}                               &\multicolumn{2}{|c|}{\underline{G191B2B}} \\
B & $4.70\pm0.04$ & $115.70\pm 0.22$ & $4.66\pm 0.07$&$ 115.5 \pm 0.2$ &\ \ \ \ \ 0.072 &\ \ \ \  -\ 0.059 \\
V & $4.79\pm0.03$ & $114.93\pm 0.17$ & $4.76\pm 0.10$&$ 114.2 \pm 0.2$ &\ \ \  -\ 0.022 &\ \ \ \  -\ 0.041 \\
R & $4.53\pm0.03$ & $114.46\pm 0.17$ & $4.56\pm 0.17$&$ 114.2 \pm 0.2$ &\ \ \  -\ 0.036 &\ \ \ \ \ \ 0.027 \\
\hline
\end{tabular}
\end{table*}

For calibration of polarization angle zero point, we observed highly polarized standard stars and  the results are given in Table 1.
The measured systematic differences were applied to the program stars.  To estimate the value of instrumental polarization, a number of unpolarized standard stars have been observed. These measurements show that the instrumental polarization is  below 0.03\% in all pass-bands. In fact, the instrumental polarization of ST has been monitored since 2004 within other projects as well (e.g. Medhi et al. 2007,2008). These measurements demonstrated that it is invariable in all B,V and R pass-band.  The instrumental polarization was then applied to all measurements.

\section{Analysis and Results}
\label{sec:analysis}

\begin{figure}
\includegraphics[height=8cm,width=8cm]{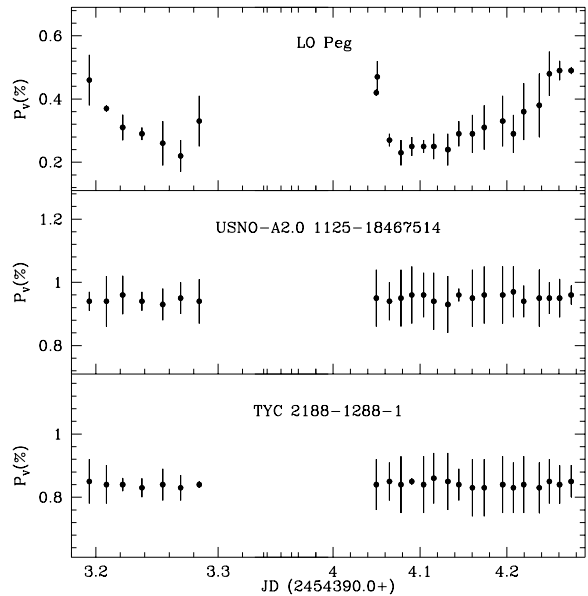}
\caption{V band polarimetric light curve of LO Peg (top panel), USNO-A2.0 1125-18467514
(middle panel) and TYC 2188-1288-1 (bottom panel)}
\label{fig:lopeg_jd}
\end{figure}

The degree of polarization  and the corresponding position angle  ~for LO Peg in each B, V and R filters are given in Table 2.  Both \p ~and \pa ~are found to  vary with time. We have plotted the degree of polarization in V band ($P_V$) against Julian day (JD) of the observations in top panel of Figure \ref{fig:lopeg_jd}, which clearly shows that \pv ~varies with JD.  In order to ascertain the polarization variability in LO Peg, we have also observed polarization of two nearby stars, namely USNO-A2.0 1125-18467514 and TYC 2128-1288-1. The results of observed degree of polarization  and position angle in V band  of these two nearby stars are given in Table 3.  The middle and bottom panels of Fig. \ref{fig:lopeg_jd} show the variation of $P_V$ against JD for the stars USNO-A2.0 1125-18467514 and TYC 2128-1288-1, respectively.  Lack of any significant variation in \p ~for these stars imply that the observed  polarization variation in LO Peg is intrinsic.  The measurements on epoch JD = 2454393.284503 show higher values ($\sim$10\degr) of position angle in comparison to the previous epoch observations, which were taken  about $22$ minutes earlier (see Table 2).  Further, the same epoch observations show a sudden decrement/increment in B, V and R bands light curves.  This effect is probably due to the observations taken at relatively high airmass. We assume that the data points at this epoch are of bad quality, therefore, are not used for further analysis.  The weighted mean of the degree of polarization and polarization position angle are calculated to be \pb ~= $0.387\pm0.004 \%$, \pv ~= $0.351\pm0.004 \%$, \pr ~= $0.335\pm0.003 \%$, and \Pab = $88\degr\pm1\degr$, \Pav ~= $91\degr\pm1\degr$, \Par ~= $91\degr\pm1\degr$.

\begin{figure}
\includegraphics[height=8cm,width=8cm]{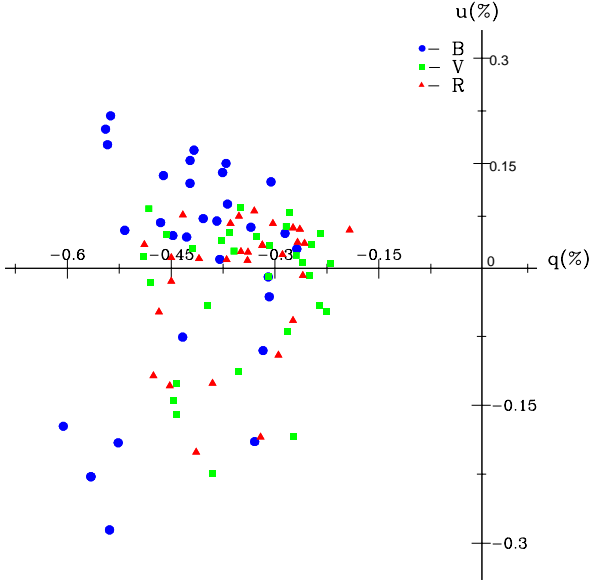}
\caption{Normalized Stokes parameters of LO Peg plotted in the ($q, u$) plane
for  B ($\bullet$), V($\blacksquare$) and R($\blacktriangle$) band.}
\label{fig:lopeg_qu}
\end{figure}

Fig. \ref{fig:lopeg_qu} depicts the movements of the polarization vector in $q$ and $u$ plane (normalized stokes parameters) for B (solid circle), V(solid square) and R (solid triangle) bands. The polarization parameters are changed due to variations in both Stokes parameters $q$ and $u$.


\begin{table*}
\caption{Results of the degree of polarization and polarization position angle observed for LO Peg.}
\begin{tabular}{cccccccccc}
\hline
\hline
JD & Phase & \multicolumn{2}{c}{B}&&\multicolumn{2}{c}{V}&&\multicolumn{2}{c}{R} \\
\cline{3-4}\cline{6-7}\cline{9-10}
   &       & $P_B$(\%) &$\theta_B$(\deg)&&$P_V$(\%) &$\theta_V$(\deg)&&$P_R$(\%) &$\theta_R$(\deg) \\
\hline
2454393.194613& 0.2529 &  $0.52\pm0.09$  &    $87 \pm5 $ &&   $0.46\pm0.08$ & $87 \pm5 $ &&   $0.49\pm0.06$  & $88 \pm4 $\\
2454393.208640& 0.2861 &  $0.44\pm0.06$  &    $95 \pm4 $ &&   $0.37\pm0.01$ & $99 \pm8 $ &&   $0.41\pm0.05$  & $89 \pm4 $\\
2454393.221914& 0.3174 &  $0.38\pm0.02$  &    $89 \pm6 $ &&   $0.31\pm0.04$ & $91 \pm3 $ &&   $0.37\pm0.02$  & $89 \pm3 $\\
2454393.237712& 0.3535 &  $0.33\pm0.07$  &    $98 \pm8 $ &&   $0.29\pm0.02$ & $97 \pm3 $ &&   $0.31\pm0.03$  & $99 \pm3 $\\
2454393.254632& 0.3945 &  $0.34\pm0.02$  &    $85 \pm2 $ &&   $0.26\pm0.07$ & $89 \pm7 $ &&   $0.29\pm0.01$  & $88 \pm2 $\\
2454393.269319& 0.4287 &  $0.31\pm0.06$  &    $93 \pm7 $ &&   $0.22\pm0.05$ & $89 \pm6 $ &&   $0.28\pm0.06$  & $90 \pm6 $\\
2454393.284503& 0.4648 &  $0.38\pm0.07$  &    $105\pm5 $ &&   $0.33\pm0.08$ & $107\pm6 $ &&   $0.37\pm0.01$  & $105\pm2 $\\
2454394.049720& 0.2705 &  $0.38\pm0.04$  &    $83 \pm4 $ &&   $0.42\pm0.01$ & $88 \pm3 $ &&   $0.35\pm0.03$  & $88 \pm2 $\\
2454394.064950& 0.3057 &  $0.27\pm0.01$  &    $87 \pm4 $ &&   $0.27\pm0.02$ & $88 \pm4 $ &&   $0.34\pm0.02$  & $88 \pm5 $\\
2454394.078097& 0.3369 &  $0.31\pm0.06$  &    $91 \pm8 $ &&   $0.23\pm0.04$ & $96 \pm4 $ &&   $0.28\pm0.03$  & $96 \pm2 $\\
2454394.090643& 0.3672 &  $0.29\pm0.01$  &    $85 \pm2 $ &&   $0.25\pm0.03$ & $91 \pm3 $ &&   $0.26\pm0.01$  & $91 \pm2 $\\
2454394.104149& 0.3994 &  $0.39\pm0.02$  &    $85 \pm3 $ &&   $0.25\pm0.02$ & $86 \pm3 $ &&   $0.26\pm0.06$  & $86 \pm4 $\\
2454394.115896& 0.4268 &  $0.43\pm0.02$  &    $87 \pm2 $ &&   $0.25\pm0.04$ & $90 \pm3 $ &&   $0.30\pm0.01$  & $90 \pm4 $\\
2454394.132122& 0.4648 &  $0.41\pm0.01$  &    $85 \pm5 $ &&   $0.24\pm0.05$ & $84 \pm3 $ &&   $0.28\pm0.02$  & $84 \pm4 $\\
2454394.144725& 0.4951 &  $0.33\pm0.01$  &    $79 \pm6 $ &&   $0.29\pm0.04$ & $82 \pm4 $ &&   $0.20\pm0.04$  & $82 \pm8 $\\
2454394.160279& 0.5312 &  $0.44\pm0.05$  &    $82 \pm5 $ &&   $0.29\pm0.06$ & $84 \pm7 $ &&   $0.31\pm0.05$  & $84 \pm5 $\\
2454394.173994& 0.5635 &  $0.45\pm0.09$  &    $79 \pm7 $ &&   $0.31\pm0.07$ & $87 \pm5 $ &&   $0.32\pm0.01$  & $87 \pm2 $\\
2454394.195231& 0.6133 &  $0.40\pm0.04$  &    $80 \pm3 $ &&   $0.33\pm0.08$ & $86 \pm7 $ &&   $0.27\pm0.05$  & $86 \pm10$\\
2454394.207499& 0.6426 &  $0.40\pm0.09$  &    $79 \pm13$ &&   $0.29\pm0.06$ & $84 \pm5 $ &&   $0.27\pm0.08$  & $84 \pm12$\\
2454394.219419& 0.6709 &  $0.47\pm0.03$  &    $86 \pm3 $ &&   $0.36\pm0.09$ & $83 \pm10$ &&   $0.34\pm0.09$  & $83 \pm13$\\
2454394.237207& 0.7129 &  $0.45\pm0.05$  &    $87 \pm9 $ &&   $0.38\pm0.10$ & $87 \pm8 $ &&   $0.32\pm0.05$  & $87 \pm4 $\\
2454394.248885& 0.7402 &  $0.58\pm0.02$  &    $80 \pm3 $ &&   $0.48\pm0.07$ & $91 \pm10$ &&   $0.45\pm0.04$  & $91 \pm3 $\\
2454394.260643& 0.7676 &  $0.57\pm0.02$  &    $81 \pm2 $ &&   $0.49\pm0.03$ & $85 \pm3 $ &&   $0.44\pm0.07$  & $85 \pm10$\\
2454394.273964& 0.7988 &  $0.58\pm0.02$  &    $79 \pm3 $ &&   $0.49\pm0.01$ & $89 \pm6 $ &&   $0.45\pm0.01$  & $89 \pm6 $\\
2454407.216597& 0.3428 &  $0.46\pm0.03$  &    $90 \pm4 $ &&   $0.24\pm0.01$ & $95 \pm3 $ &&   $0.34\pm0.01$  & $89 \pm3 $\\
2454408.203042& 0.6699 &  $0.48\pm0.02$  &    $82 \pm3 $ &&   $0.36\pm0.03$ & $88\pm3 $ &&   $0.36\pm0.02$  & $84 \pm2 $\\
2454421.047389& 0.9814 &  $0.63\pm0.05$  &    $98 \pm3 $ &&   $0.46\pm0.08$ & $98 \pm5 $ &&   $0.47\pm0.04$  & $93 \pm2 $\\
2454421.058511& 0.0078 &  $0.61\pm0.07$  &    $104\pm6 $ &&   $0.45\pm0.04$ & $105\pm5 $ &&   $0.46\pm0.05$  & $103\pm3 $\\
2454421.103554& 0.1133 &  $0.61\pm0.03$  &    $101\pm2 $ &&   $0.47\pm0.04$ & $100\pm3 $ &&   $0.49\pm0.09$  & $97 \pm6 $\\
2454421.132452& 0.1826 &  $0.56\pm0.05$  &    $100\pm3 $ &&   $0.40\pm0.05$ & $93 \pm3 $ &&   $0.41\pm0.06$  & $99 \pm4 $\\
2454422.165401& 0.6201 &  $0.45\pm0.04$  &    $80 \pm3 $ &&   $0.37\pm0.03$ & $86 \pm5 $ &&   $0.37\pm0.03$  & $85 \pm3 $\\
2454454.050914& 0.8662 &  $0.61\pm0.04$  &    $101\pm2 $ &&   $0.47\pm0.05$ & $99 \pm3 $ &&   $0.47\pm0.06$  & $98 \pm3 $\\
\hline
\end{tabular}
\end{table*}

\begin{table}
\caption{ V band polarization and position angle of USNO-A2.0~1125-18467514 (star1) and TYC 2188-1288-1 (star2).}
\label{tab:comp_pol}
\begin{tabular}{cccccc}
\hline
\hline
JD   &\multicolumn{2}{p{0.5in}}{ star1} &&\multicolumn{2}{p{0.5in}}{star2} \\
\cline{2-3} \cline{5-6}
2454000.0+     &     $P_V(\%)$      &  $\theta_V(\deg)$ &&  $P_V(\%) $     &   $\theta_V(\deg)$      \\
\hline
  393.194613  & $0.94\pm0.03$  & $ 99\pm2 $   &&    $0.85\pm0.07$   &    $ 98\pm10$\\
  393.208640  & $0.94\pm0.08$  & $100\pm3 $   &&    $0.84\pm0.06$   &    $107\pm6 $\\
  393.221914  & $0.96\pm0.06$  & $ 99\pm3 $   &&    $0.84\pm0.02$   &    $107\pm2 $\\
  393.237712  & $0.94\pm0.03$  & $102\pm2 $   &&    $0.83\pm0.03$   &    $104\pm9 $\\
  393.254632  & $0.93\pm0.05$  & $ 97\pm5 $   &&    $0.84\pm0.05$   &    $ 99\pm8 $\\
  393.269319  & $0.95\pm0.05$  & $104\pm5 $   &&    $0.83\pm0.04$   &    $ 96\pm7 $\\
  393.284503  & $0.94\pm0.07$  & $100\pm3 $   &&    $0.84\pm0.01$   &    $ 98\pm2 $\\
  394.049720  & $0.95\pm0.09$  & $ 99\pm4 $   &&    $0.84\pm0.08$   &    $ 96\pm9 $\\
  394.064950  & $0.94\pm0.06$  & $ 98\pm8 $   &&    $0.85\pm0.06$   &    $101\pm6 $\\
  394.078097  & $0.95\pm0.09$  & $100\pm7 $   &&    $0.84\pm0.09$   &    $ 94\pm8 $\\
  394.090643  & $0.96\pm0.09$  & $ 90\pm8 $   &&    $0.85\pm0.01$   &    $ 88\pm2 $\\
  394.104149  & $0.96\pm0.07$  & $101\pm3 $   &&    $0.84\pm0.09$   &    $100\pm11$\\
  394.115896  & $0.94\pm0.09$  & $ 97\pm12$   &&    $0.86\pm0.08$   &    $103\pm6 $\\
  394.132122  & $0.93\pm0.09$  & $ 98\pm8 $   &&    $0.85\pm0.09$   &    $ 97\pm5 $\\
  394.144725  & $0.96\pm0.02$  & $ 95\pm2 $   &&    $0.84\pm0.05$   &    $ 94\pm2 $\\
  394.160279  & $0.95\pm0.09$  & $106\pm5 $   &&    $0.83\pm0.09$   &    $107\pm9 $\\
  394.173994  & $0.96\pm0.09$  & $108\pm8 $   &&    $0.83\pm0.09$   &    $104\pm4 $\\
  394.195231  & $0.96\pm0.09$  & $ 97\pm12$   &&    $0.84\pm0.09$   &    $102\pm5 $\\
  394.207499  & $0.97\pm0.08$  & $ 96\pm11$   &&    $0.83\pm0.08$   &    $ 98\pm7 $\\
  394.219419  & $0.94\pm0.05$  & $109\pm3 $   &&    $0.84\pm0.09$   &    $ 93\pm5 $\\
  394.237207  & $0.95\pm0.09$  & $ 99\pm5 $   &&    $0.83\pm0.08$   &    $ 90\pm5 $\\
  394.248885  & $0.95\pm0.05$  & $ 92\pm6 $   &&    $0.85\pm0.07$   &    $ 94\pm6 $\\
  394.260643  & $0.95\pm0.06$  & $ 99\pm3 $   &&    $0.84\pm0.06$   &    $ 97\pm8 $\\
  394.273964  & $0.96\pm0.03$  & $102\pm2 $   &&    $0.85\pm0.05$   &    $ 98\pm3 $\\
  407.216597  & $0.93\pm0.05$  & $ 83\pm2 $   &&    $0.84\pm0.05$   &    $ 85\pm2 $\\
  408.203042  & $0.96\pm0.07$  & $101\pm3 $   &&    $0.85\pm0.09$   &    $ 97\pm5 $\\
  421.047389  & $0.94\pm0.08$  & $ 96\pm7 $   &&    $0.84\pm0.08$   &    $101\pm7 $\\
  421.058511  & $0.94\pm0.06$  & $104\pm2 $   &&    $0.85\pm0.06$   &    $107\pm2 $\\
\hline
\end{tabular}
\end{table}

LO Peg is located at a distance of 25 pc and therefore has a negligible reddening.  It is quite natural to assume that the observed polarization in LO Peg is not  foreground in origin. Further, the observed polarization in any distant star located near the Galactic plane may have small negligible interstellar component.  We have observed polarization of 17 field stars within a 8\arcm ~radius around LO Peg. The results of V-band polarization are given in Table 4. The average values of the \pv ~and \pav ~computed from the data obtained for the field stars are found to be 1.84\%  and 94\deg, respectively.

\begin{table}
\caption{ V band polarization and the position angle of the stars lying within the radius of  8\arcm ~from LO Peg.}
\label{tab:filed_pol}
\begin{tabular}{rcccc}
\hline
\hline
ID & USNO A2- No.  &  B(mag) & $P_V$ (\%)      & $\theta_V$ (\deg) \\
\hline
1  &  1125-18467514& 12.5    &   $0.76\pm0.08$ & $ 98\pm  9 $\\
2  &  1125-18470176& 14.5    &   $1.16\pm0.09$ & $ 96\pm  3 $\\
3  &  1125-18467687& 15.5    &   $1.96\pm0.07$ & $ 86\pm  3 $\\
4  &  1125-18466948& 14.0    &   $0.59\pm0.04$ & $ 85\pm  2 $\\
5  &  1125-18465461& 14.5    &   $2.19\pm0.09$ & $ 47\pm  2 $\\
6  &  1125-18463910& 14.6    &   $2.97\pm0.09$ & $178\pm  2 $\\
7  &  1125-18463552& 15.5    &   $3.21\pm0.19$ & $140\pm  2 $\\
8  &  1125-18466649& 15.4    &   $1.02\pm0.05$ & $ 87\pm  2 $\\
9  &  1125-18467076& 14.8    &   $0.71\pm0.07$ & $ 16\pm  3 $\\
10 &  1125-18472889& 13.5    &   $0.65\pm0.10$ & $ 98\pm  5 $\\
11 &  1125-18471384& 12.6    &   $0.67\pm0.09$ & $101\pm  7 $\\
12 &  1125-18470313& 14.7    &   $1.52\pm0.09$ & $ 98\pm  5 $\\
13 &  1125-18466610& 15.5    &   $3.40\pm0.25$ & $ 84\pm  2 $\\
14 &  1125-18468115& 15.3    &   $3.46\pm0.02$ & $102\pm  2 $\\
15 &  1125-18469000& 15.8    &   $0.74\pm0.06$ & $114\pm  3 $\\
16 &  1125-18473126& 15.6    &   $1.81\pm0.07$ & $112\pm  2 $\\
17 &  1125-18472836& 15.9    &   $3.27\pm0.22$ & $101\pm  3 $\\
\hline
\end{tabular}
\end{table}

\subsection{Phase-locked polarimetric variability}

\begin{figure*}
\centering
\subfigure[]{\includegraphics[width=7cm,height=12cm]{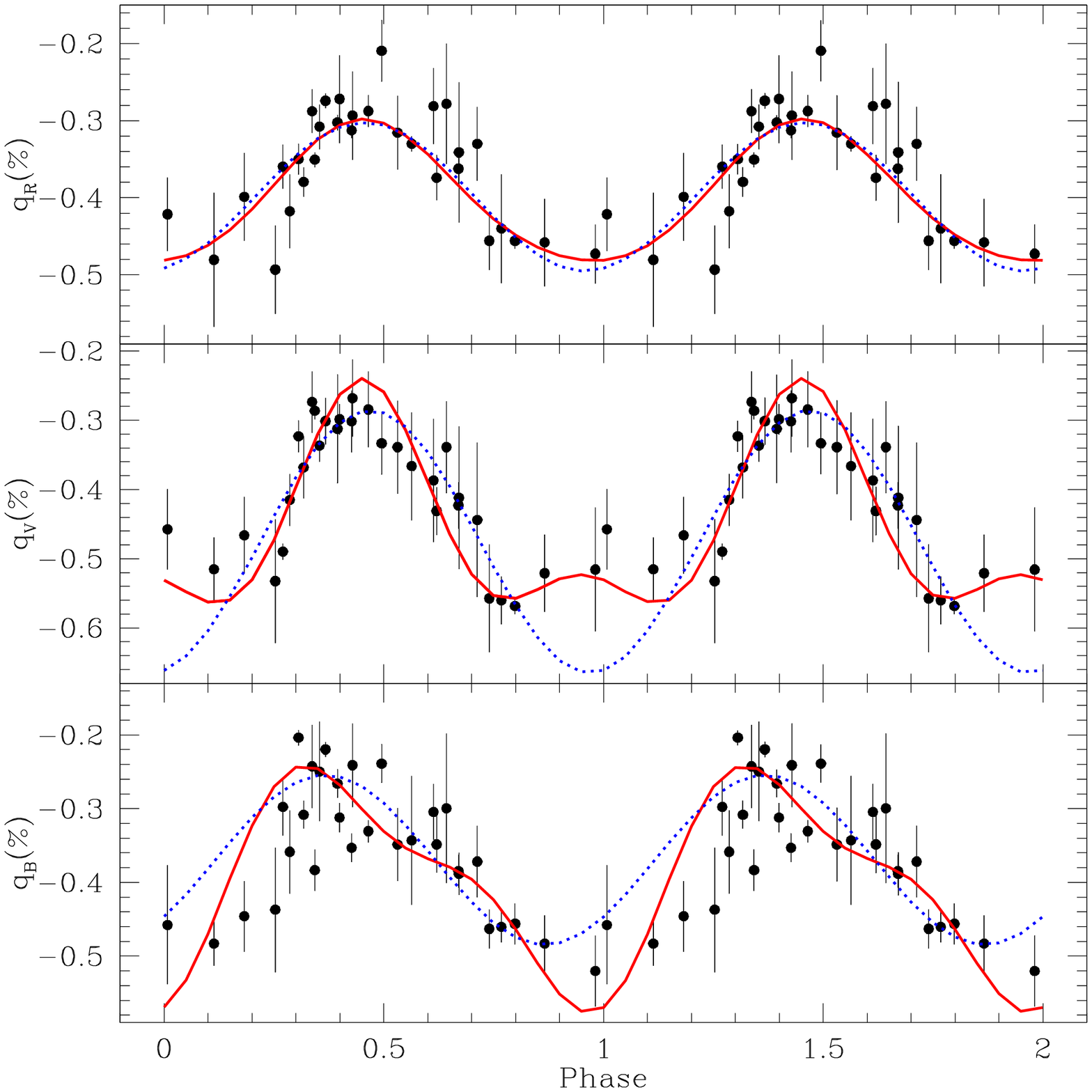}}
\subfigure[]{\includegraphics[width=7cm,height=12cm]{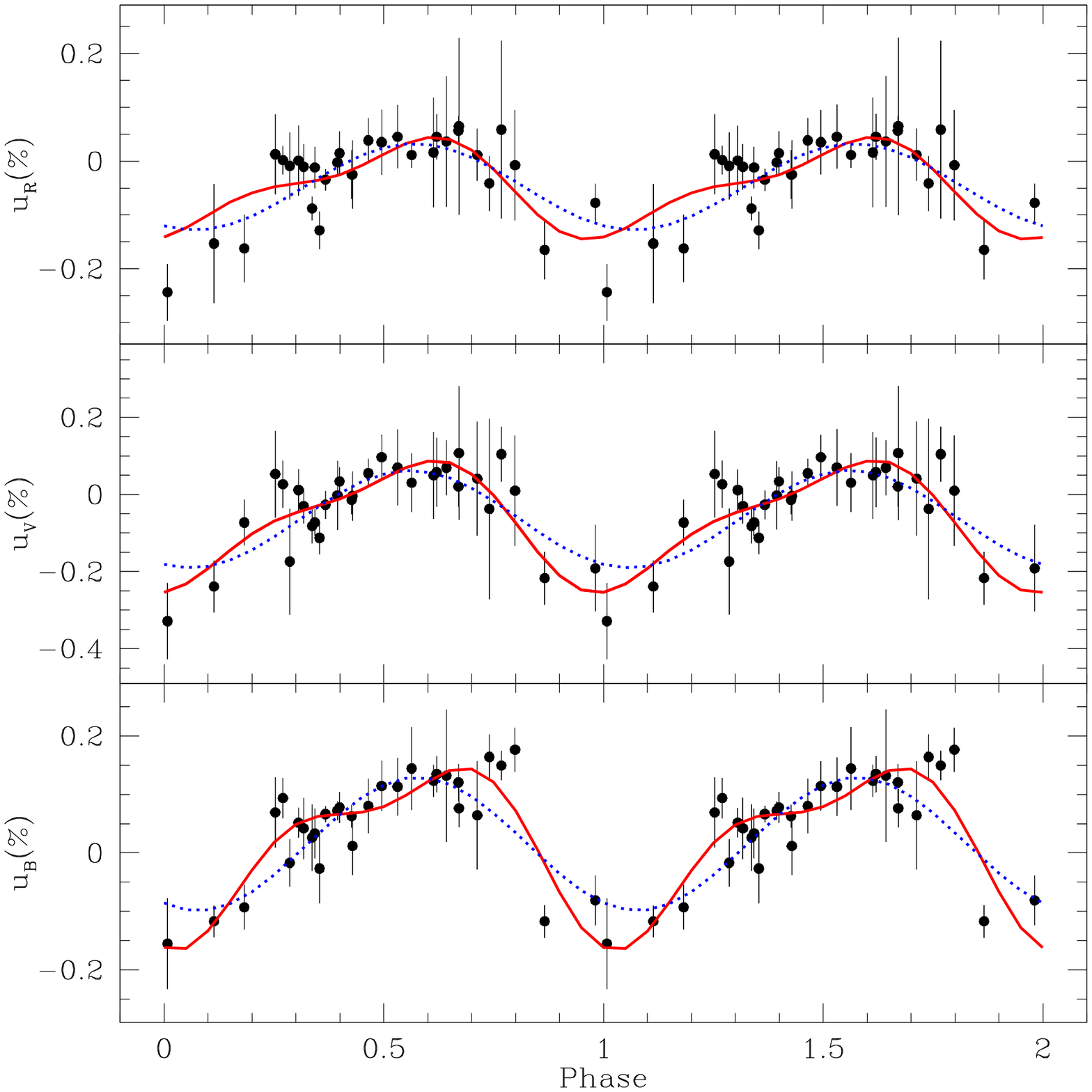}}
\caption{B, V and R polarimetry of LO Peg expressed in terms of the normalised
Stokes  parameters $q = P cos2\theta$ and $u = P sin2\theta$, plotted against
rotational phase.  The dotted and continuous lines are the weighted first and second order Fourier fits,
respectively.}
\label{fig:lopeg_ph}
\end{figure*}

The variability of the linear polarization  in LO Peg exhibits a distinct dependence on the rotational phase. Therefore, to see the phase locked polarimetric variability, we have folded the polarization data using the ephemeris HJD = 2,448,869.93 + 0.42375 E provided by dal \& Tas 2003. Figs. \ref{fig:lopeg_ph} (a) and (b) show the plots of normalised Stokes vectors q and u in B, V and R bands as a function of  photometric phase, respectively. It appears that both q and u vectors  depend on the stellar longitude.  Light scattering by circumstellar material in LO Peg could give rise to harmonic variations in observed polarization with rotational period.  Fourier fitting of the observational data with the orbital phase ($\phi$) is commonly used to analyze the polarization variability in binaries (e.g. Berdyugin et al. 2006). However, this type of analysis is useful for polarimetric variability in single spotted star, because,  for example, a single spot can give power in the second harmonics  due to scattering phase function.

\begin{equation}
P_\lambda(\phi) = A + \sum_{k=1}^{2} ~B_k ~{\rm cos}(k2\pi\phi) + \sum_{k=1}^{2} ~C_k ~{\rm sin}(k2\pi\phi)
\end{equation}

\noindent
where, $\phi$ is rotational phase.  Firstly, a first order Fourier series was fitted to the polarimetric waves. To see the presence of second harmonic (if any) in the polarimetric  waves, later we fit second order Fourier series to the data.  Results of these fits with reduced $\chi^2$ ($\chi^2_\nu$) are given in Table 5.  The best fits with the polarimetric data are shown in Fig. \ref{fig:lopeg_ph}, where dotted and continuous lines are the weighted  first  and second order Fourier fits, respectively.   Fig. \ref{fig:lopeg_ph} and Table 5 clearly show that  the fit of second order Fourier series improves the $\chi^2_\nu$ for B and V filters.  However, for R filter no significant improvement was found in the $\chi^2_\nu$ while fitting the data with second order Fourier series.  The semi-amplitudes of the polarimetric waves are determined by using the relation $\Delta P = \sqrt{(B_1^2 +C_1^2)}$.  The semi-amplitudes of q and u waves are thus determined to be $0.12\pm$0.03 and $0.17\pm0.03$, $0.10\pm0.03$ and $0.09\pm0.02$, and $0.10\pm0.02$ and $0.07\pm0.02$ in B, V and R bands, respectively.

\begin{figure*}
\centering
\includegraphics[width=14cm,height=16cm]{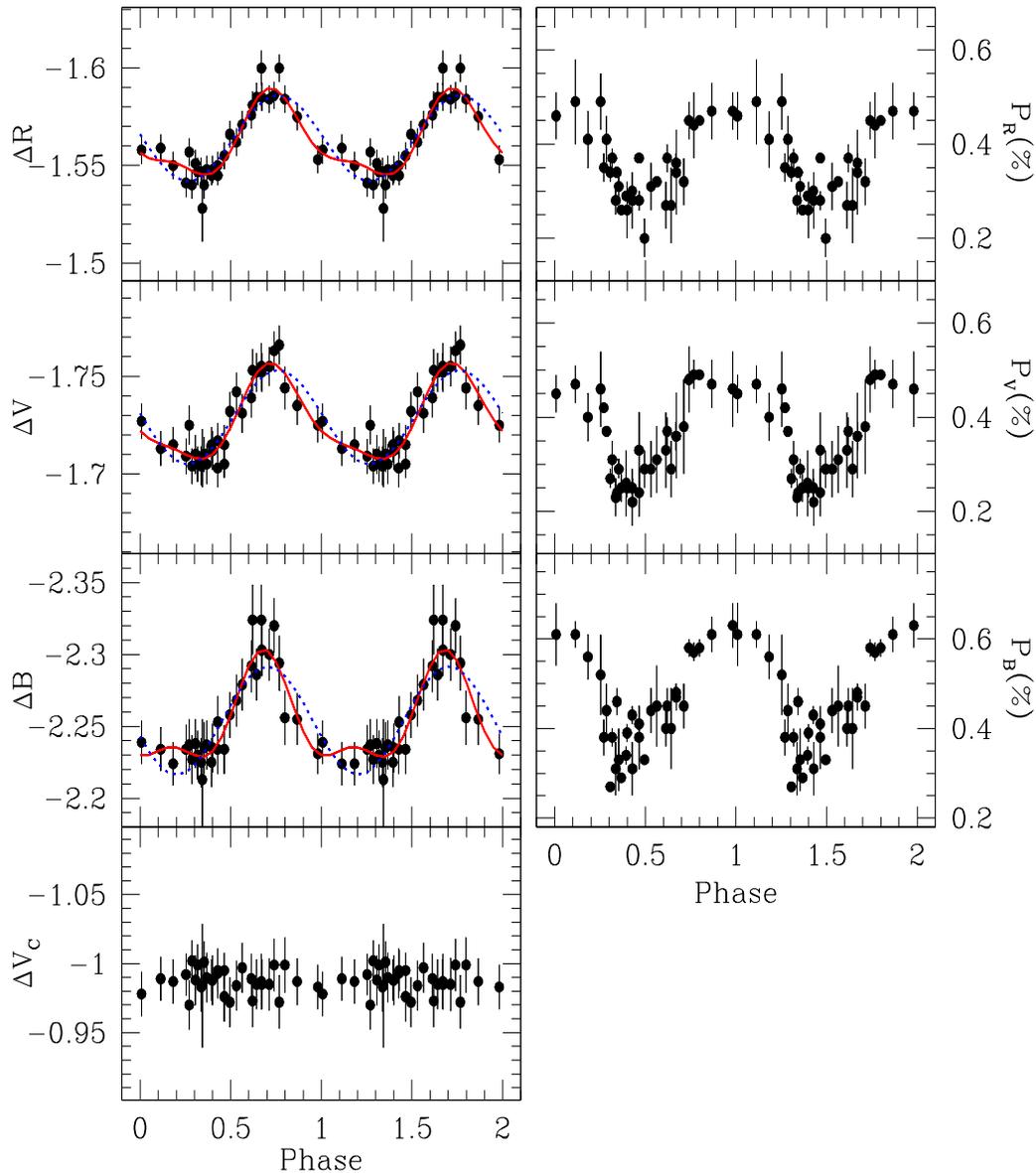}
\caption{Left panel: Three upper plots show B, V and R band light curves of LO Peg folded using the period 0.42375 d. The lowermost plot shows the differential light curve of comparison star with respect to check star,  right panel: B, V and R polarimetric light curve of LO Peg as a function of photometric phase. The dotted and continuous lines in the right panels are the weighted first and second order Fourier fits, respectively.}
\label{fig:lopeg_lc}
\end{figure*}

\begin{table*}
\caption{Results from Fourier fit.}
\begin{tabular}{ccccccc}
\hline
\hline
Filter &$A            $ & $B_1           $ & $ C_1         $ &  $ B_2          $ & $C_2             $ & $\chi^2_\nu$/dof \\
\hline
&&&&&&\\
\multicolumn{7}{c}{q}\\

B&$-0.445\pm0.011$& $ 0.091\pm0.011$& $-0.082\pm0.018$&  $              $& $               $& 2.37/28 \\
 &$-0.476\pm0.012$& $ 0.082\pm0.012$& $-0.127\pm0.019$&  $-0.005\pm0.014$& $ -0.055\pm0.013$& 2.04/26 \\
V&$-0.406\pm0.007$& $ 0.033\pm0.008$& $-0.167\pm0.015$&  $              $& $               $& 1.84/28 \\
 &$-0.387\pm0.010$& $ 0.036\pm0.007$& $-0.122\pm0.015$&  $-0.038\pm0.009$& $ 0.053\pm0.011 $& 1.49/26 \\
R&$-0.390\pm0.005$& $ 0.027\pm0.005$& $-0.097\pm0.008$&  $              $& $               $& 1.44/28 \\
 &$-0.390\pm0.009$& $ 0.023\pm0.008$& $-0.093\pm0.013$&  $-0.007\pm0.008$& $ 0.007\pm0.008 $& 1.44/26 \\
&&&&&&\\
\multicolumn{7}{c}{u}\\
B&$-0.007\pm0.012$& $-0.080\pm0.013$& $-0.153\pm0.018$&  $              $& $               $& 1.57/28 \\
 &$-0.029\pm0.010$& $-0.078\pm0.012$& $-0.183\pm0.016$&  $-0.004\pm0.012$& $-0.084\pm0.01  $& 1.26/26 \\
V&$-0.035\pm0.008$& $-0.033\pm0.008$& $-0.084\pm0.011$&  $              $& $               $& 0.99/28 \\
 &$-0.040\pm0.007$& $-0.024\pm0.010$& $-0.106\pm0.011$&  $ 0.022\pm0.009$& $-0.025\pm0.009 $& 0.90/26 \\
R&$-0.021\pm0.007$& $-0.031\pm0.008$& $-0.067\pm0.011$&  $              $& $               $& 1.17/28 \\
 &$-0.021\pm0.007$& $-0.015\pm0.012$& $-0.071\pm0.010$&  $ 0.024\pm0.011$& $ -0.015\pm0.008$& 1.11/26 \\
&&&&&&\\
\multicolumn{7}{c}{Light Curve}\\
B&$-2.254\pm0.003$& $ 0.036\pm0.004$& $ 0.011\pm0.004$&  $              $& $               $& 0.84/28 \\
 &$-2.254\pm0.002$& $ 0.031\pm0.003$& $ 0.014\pm0.003$&  $-0.012\pm0.003$& $0.010\pm0.002  $& 0.60/26 \\
V&$-1.729\pm0.002$& $ 0.024\pm0.002$& $-0.002\pm0.002$&  $              $& $               $& 0.77/28 \\
 &$-1.738\pm0.001$& $ 0.023\pm0.002$& $ 0.001\pm0.002$&  $-0.004\pm0.002$& $0.005\pm0.002  $& 0.65/26\\
R&$-1.564\pm0.002$& $ 0.022\pm0.002$& $-0.002\pm0.002$&  $              $& $               $& 1.03/28 \\
 &$-1.563\pm0.001$& $ 0.020\pm0.002$& $ 0.0005\pm0.002$& $-0.004\pm0.002$& $ 0.006\pm0.002 $& 0.80/26 \\
\hline
\multicolumn{7}{l}{{\it Note:} dof is degree of freedom}\\
\end{tabular}
\end{table*}

\subsection{Photometric Variability}
Photometric light curves in B, V and R bands are obtained from the polarimetric data.  The stars USNO-A2.0~1125-18467514 and TYC 2188-1288-1 were used as comparison and check stars, respectively.  Differential photometry in the sense of subtracting the comparison from the variable was done, as all the program, comparison, and check stars were in the same CCD frame.  Variations in the B, V and R magnitudes ($\Delta$B, $\Delta$V and $\Delta$R) of the program star are shown in left panel of Fig. \ref{fig:lopeg_lc}. The different measures of comparison and check stars ($\Delta V_c$) is shown  in the bottom plot of left panel of Fig. \ref{fig:lopeg_lc}. The standard deviation ($\sigma$) of the different measures of the comparison and check stars were determined to be 0.012 in  B, 0.008 in V and 0.012 in R filters. This implies that the comparison star was constant during the observations.   The phase of minima  was determined by a linear least-square fitting of the second-order polynomial at the minimum of light curve. The minima was thus determined to be $0.27\pm0.05$, $0.30\pm0.05$ and $0.30\pm0.05$ in B, V and R bands, respectively.  Recently, Pandey et al. (2005) found that the phase minima in the light curve of the star LO Peg shifts with a rate of 0.85\deg ~per day. Using this value of the phase shift and the mean epoch of present polarimetric observations (JD = 2454403.37), the minimum of the phase was computed to be 0.3, which is similar to that obtained from the photometric light curve. We have also fitted the first and  second order Fourier series to the photometric light curves and these are shown by dotted and continuous lines, respectively in Fig. \ref{fig:lopeg_lc}. The results of these fits are given in Table 5. The second order Fourier fit over the first order Fourier fit  to these data reduces the $\chi_\nu^2$ by 20\% at least.  This implies that the photometric variability is probably due to two inhomogeneous regions on the surface of the star.  The photometric amplitude in B , V and R bands are determined to be 0.08, 0.05 and 0.05 mag, respectively.

To compare the photometric variations with that of polarimetric variations, we have plotted degree of polarization as a function of photometric phase in the right panel of Fig. \ref{fig:lopeg_lc}. The dependence of polarization on the stellar longitude is clearly seen.  The minimum of polarimetric light curve was determined to be 0.37, 0.40 and 0.40 in B, V and R filters, respectively.  These values of minima are slightly more than that of the corresponding photometric phase minima of the same epoch.  This implies that variation in the degree of polarization of the star appears to be correlated with its photometric brightness.  The significance of correlation has been calculated by determining the linear correlation coefficient r between the magnitude and the degree of polarization. The value of r between $\Delta B$ and \pb, $\Delta V$ and \pv, and  $\Delta R$ and \pr ~was found to be -0.28, -0.47 and -0.16 with the corresponding probability of no correlation being 0.013, 0.008, and 0.043, respectively. The semi-amplitude of polarimetric waves are determined to be $0.18\pm0.02$\%, $0.13\pm0.01$\% and $0.10\pm0.02$\% in B, V and R bands, respectively.

\section{Discussion and Conclusions}
\label{sec:discussion}
We have carried out analysis of the B, V and R bands polarimetric observations of the single late-type active dwarf LO Peg.  A polarization upto 0.63\% in B band was observed.  Our analysis shows that  polarization in LO Peg is intrinsic rather than having an interstellar origin. Further, the existence of a time dependent polarization is a well established criterion for intrinsic polarization (Zellner \& Serkowski 1972).  Such a large polarization was also observed for the active dwarfs MS Ser (0.4\% in U band; Alekseev 2003), LQ Hya (0.28\% in U band; Alekseev 2003), TU Pyx (0.5\% in V band; Clarke, Smith \& Yudin 1998). However, polarization in LO Peg  was found to be much more than that of some solar-type (G-type) stars, where observed polarizations were  of the order of  $10^{-2}$\% (Leroy \& Le Borgne 1989).  The observed polarization  in LO Peg was found to decrease toward longer wavelength.  Similar trend has been observed for the  most of active stars (e.g. Yudin \& Evans 2002, Rostopchina et al. 2007).  This could be due to number of factors. For example;  i) selective absorption by circumstellar dust, which grows toward shorter wavelengths. As a result, the ratio of the intensity of the scattered (polarized) light to the intensity of star's light coming directly to us (unpolarized) increases with decreasing wavelength of the radiation,  ii) wavelength dependent  albedo, which decreases toward longer wavelengths resulting less scattering and thus polarization.  The mechanism  of the polarization in these stars are best interpreted by magnetic intensification or scattering by circumstellar material.   Below we discuss on the origin of polarization.

\begin{figure}
\centering
\includegraphics[width=6cm,height=8cm,angle=-90]{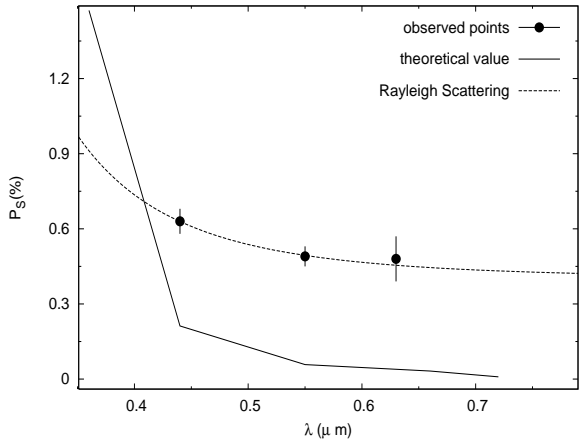}
\caption{Wavelength dependence of the degree of polarization. The observed data points
are represented by solid dots and solid curve shows the expected theoretical values.
Dashed line shows the Rayleigh scattering}
\label{fig:model}
\end{figure}

Magnetic polarization has been modeled in a number of studies and degree of polarization is assumed to depend linearly on the size of magnetized region (e.g. Degl'Innocenti 1983, Calamai \& Degl'Innocenti 1983, Mullan \& Bell 1976).  However, later on it was demonstrated that this dependence becomes nonlinear for large regions (see Huovelin \& Saar 1991,  Saar \& Huovelin 1993). In particular, for a single circular region the degree of polarization will be proportional to a factor $A$, which approximately depends on the area of the region ($f$ in \%) as (Saar \& Huovelin 1993):

\begin{eqnarray}
A(f)& = &\frac{P_{S}(f)}{P_{S}(f=1\%)}\nonumber\\
    & \approx &-2.128\times10^{-4} +1.076f-4.812 f^{2} + 9.058f^3 \nonumber \\
    &&- 6.26f^4
\end{eqnarray}

\noindent
Saar \& Huovelin (1993) have also calculated a grid of expected degrees of polarization in UBVRI bands for the  stars having temperatures from 4000 to 7000 \degr K and log $g$ from 2.0 to 4.5. We have used their results to compare with our observed values of maximum polarization of LO Peg. The maximum values of the degree of polarization ($P_S$) in B, V and R bands for the LO Peg were  found to be $0.63\pm0.05$ \%, $0.49\pm0.05$ \% and $0.48\pm0.09$ \%, respectively. Fig. \ref{fig:model} represents the maximum degree of polarization for LO Peg in B, V and R bands. The maximum possible degree of polarization for total spot area of $f \approx 24$ \% is derived from the calculations of Saar \& Huovelin (1993) for a star corresponding to the spectral type of K7V and characteristic magnetic field of 2.7 kG. These values are over-plotted as solid line in Fig.  \ref{fig:model}. The maximum observed B, V and R bands polarization exceed the theoretical values expected for Zeeman polarization model. This is probably due the presence of supplementary source of linear polarization, such as remnant of circumstellar disk or envelope, which produce the linear broad band polarization.

The present polarimetric observations of LO Peg show a clear variation in q and u.  The maximum variability in the polarization was observed in B band. Variations upto 0.7\%  in U band for a K dwarf MS Ser (Alekseev 2003, 2000), 0.1\% in B and V bands for F9 dwarf 59 Vir (Clarke, Smith \& Yudin 1998) were also observed. The variations in the polarization are appear to be correlated with the photometric variations (i.e variation in \p ~per cent followed closely the variations in brightness, in the sense that, when the star was brighter \p ~was higher).  This implies that the surface brightness inhomogeneity seems to play a major role in the polarization variability of LO Peg.  Inhomogeneities in the distribution of  both the material in the circumstellar envelope and brightness on the surface of the illuminating star can provide the necessary asymmetry, and hence produces a net polarization in the integrated light (Menard \& Bastein 1992). The dark spots and the possible associated plage regions produce the variation in the relative fluxes of the radiation in different bands across the stellar surface, which thereby produce  the variation in observed polarization.  The mechanism which can produce periodic linear polarization variability is Rayleigh or Thomson scattering in a non-spherically symmetric distribution of circumstellar material. This has been discussed by Brown \& Mclean (1977), and Simmons (1982) for a single star with a distortion produced by rotation. More recently, Al-Malki et al. (1999) have discussed the scattering polarization arising from  light source anisotropy.  However, the absence of near IR (Pandey et al. 2005) and mid IR (Chen et al. 2005) excess indicates that no dusty disk is present upto a distance of $\sim 30$ AU around LO Peg.  At this distance the scattered radiation will be diluted which results a low level of the  polarization.  The polarization arising from the scattering by an equatorial thin disk with an inclination angle of 70\degr ~is theoretically calculated to be  $<0.2$\% (e. g. Hoffman et al. 2003), which is less than that observed in LO Peg. Al-Malki et al. (1999) showed that a $\sim 0.5\%$ of polarization can be achieved from an anisotropic light source scattered by a Maclaurian spheroidal envelope (a=b=1 and c$\approx$ 0.6) with an optical depth of 0.1 and  an angle of inclination of 60\degr.  The axial inclination of LO Peg is $50\pm10$\degr (Eibe et al. 1999).

The other possible region for the polarization of LO Peg is presence of the solar prominence like structures, which are made of the condensed cool matter suspended in the corona by magnetic fields.  Cool clouds of neutral material forced to co-rotate with underlying star have been detected in the fast rotating K dwarf AB Dor, several G dwarfs in $\alpha$ Per,  and the active M dwarf  HK Aqr and RE 1816+541 (Collier-Cameron \& Robinson 1989a,b; Collier-Cameron \& Woods 1992; Byrne, Eibe \& Rolleston 1996; Eibe 1998). The characteristic dimensions and heights of these structures are considerably larger than solar prominences.  The lifetimes and growth time-scales of these clouds varies from few days to few months (e.g. see Pfeiffer 1979; Jeffries 1993).  Recent optical spectroscopic observations of LO Peg by Eibe et al. 1999 have suggested the evidence of downflow of material. They also suggested that the possibility of the condensations of cool prominence cloud can not be ruled out.  If  most of the scatterer  were gas atoms or molecules on the clouds then Rayleigh scattering from the star  dominates in contributing the polarization (see Fig. \ref{fig:model} ).  Assuming that the size of the cool prominence cloud is small compared with  the distance from the star then the polarization is given by (Pfeiffer 1979):

 \begin{equation}
 P = \frac{N \sigma f}{r_c^2 + N \sigma f}
 \label{eq:plam}
 \end{equation}

\noindent
For Rayleigh scattering

\begin{equation}
\sigma f = (8\pi e^4/3m_e^2 c^4)(\lambda_0/\lambda)^4[3(cos^2\phi +1)/16\pi]
\end{equation}

\noindent
where $f$ is the angular scattering function, $\sigma$ is the scattering cross section, $N$ is the number of scatterer, $r_c$ is the distance of the cloud from the star and $\phi$ is the scattering angle. If we assume that cool prominence cloud contain mostly hydrogen in the ground state, the resonant  wave length $\lambda_0$ is 0.122 $\mu m$.  Thus, for \pv ~= 0.35\% and assuming the cloud is located at Keplerian co-rotation radius ($r_c = (GM_\star/\Omega^2)^{1/3} = 2.83R_\star$; $M_\star$ and $R_\star$ are stellar mass and radius, respectively, $\Omega = p/2\pi$ and $p$ is rotational period) and scattered radiation received from the circumstellar cloud is completely polarized ($\phi = 90^{\circ}$), the equation (\ref{eq:plam}) yields $N = 1.1\times10^{48}$.  If the cloud is mostly hydrogen, this amounts to about $1.7\times10^{24}$ gm, which is six to seven order more than the estimated mass of the cloud of stellar prominences and $\sim$ ten order more than the Solar quiescent prominence (Collier-Cameron et al. 1990; Collier-Cameron 1996; Low, Fong \& Fan 2003 ). However, this value of mass is only an order larger than the mass loss rate  ($2.0\times10^{-11}$ M$_\odot$/year; Chen et al. 2005) observed for LO Peg.  Each particle in the cloud can be considered to have an effective cross-sectional area of $\sigma f$ as seen by the observer. Therefore, the minimum projected area of the cloud onto the plane of the sky is $N\sigma f$.  This leads the minimum value of the area of the order of $10^{21}$ cm$^2$. This value of area  is comparable to the stellar and solar prominence areas (Collier-Cameron 1996).  The estimated mass is too large to  be produce by single solar-type prominence.  This value would be an upper limit if all the scattering atoms in the star are not located in this compact, spherical cloud.  Further, N would be considerably less if scatterers were closer to the star than the assumed distance.

\begin{figure}
\centering
\includegraphics[width=6cm,height=8cm,angle=-90]{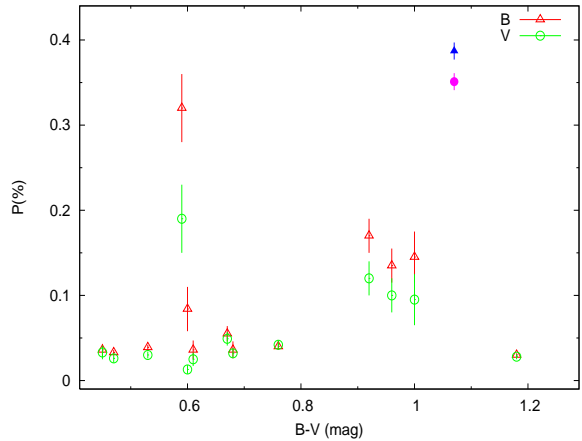}
\caption{The average values of polarization (\p) in B and V bands vs. (B-V) colour. The polarization of LO Peg in B and V bands are represented by solid circle and solid triangle, respectively. }
\label{fig:bv_pol}
\end{figure}

The average values of the polarization position angles were found similar in   B, V  and R bands.  This indicates that the scattering geometry is identical at all wavelengths. We have plotted the average values of polarization (P) of  dwarfs against (B-V) colour to see the possible spectral type dependence of polarimetric variations. The polarization values of other dwarf spectral classes  are taken from Huovelin et al. (1988) and Alekseev (2003). The distribution in the B and V bands indicates increasing linear polarization towards later spectral type. Similar trend was also observed by Huovelin et al. (1988) in U band for main sequence stars. The values of linear polarization in B and V band for LO Peg are also shown in the Fig. \ref{fig:bv_pol}. It appears that the observed  values of linear polarization for LO Peg follow the trend seen for the main-sequence stars.

If we compare the present photometric light curve of LO Peg with previous observations by Pandey et al. (2005), a shift in the phase of the minimum and a variable amplitude are quite evident. This implies that photometric variation is due to the inhomogeneities on the surface of the star. The best fit second order Fourier series indicates that the presence of two inhomogeneous region on the surface of the star. Similar signature was also reported by Pandey et al. (2005).

Therefore, we conclude that high values of  polarization observed in LO Peg require either a spheroidal envelope with an optical depth of 0.1 or clumpy material (e. g. solar prominence like structures) of mass of the order of $\sim 10^{-10}$ M$_\odot$.
\section*{Acknowledgment}
We thank our referee for his/her highly valuable comments and suggestions.  The authors also thank Prof. H. C. Bhatt for useful discussion.


\end{document}